\documentclass [prb,reprint,notitlepage,nofootinbib,superscriptaddress,floatfix] {revtex4-2}
\usepackage{physics}
\usepackage{amssymb,amsmath,mathdots}
\usepackage{graphicx}
\usepackage{lmodern}
\usepackage{amsmath}
\usepackage{bm}
\usepackage[dvipsnames]{xcolor}
\usepackage[colorlinks=true]{hyperref}
\hypersetup{
    colorlinks=true,
    linkcolor=blue,
    citecolor=blue,
    urlcolor=blue
} 
\usepackage{color}

\newcommand{\beq} {\begin{equation}}
\newcommand{\eeq} {\end{equation}}
\newcommand{\bea} {\begin{eqnarray}}
\newcommand{\eea} {\end{eqnarray}}
\newcommand{\be} {\begin{equation}}
\newcommand{\ee} {\end{equation}}
\renewcommand{\(}{\left(}
\renewcommand{\)}{\right)}
\renewcommand{\[}{\left[}
\renewcommand{\]}{\right]}
\newcommand{\Langle}{\left\langle}
\newcommand{\Rangle}{\right\rangle}

\newcommand{\non}{\nonumber}
\newcommand{\adj}[1]{{\rm ad}_{#1}}
\newcommand{\coadj}[1]{{\rm ad}^{*}_{#1}}
\newcommand{\Adj}[1]{{\rm Ad}_{#1}}
\newcommand{\coAdj}[1]{{\rm Ad}^{*}_{#1}}

\newcommand{\ve}[1]{{\bm #1}}
\newcommand{\vp}{\ve{p}}
\newcommand{\vq}{\ve{q}}
\newcommand{\vx}{\ve{x}}
\newcommand{\vy}{\ve{y}}
\newcommand{\vk}{\ve{k}}
\newcommand{\vR}{\ve{R}}
\newcommand{\Rhat}{\hat{\ve{R}}}

\newcommand{\iu}{\mathrm{i}} 
\newcommand{\eu}{\mathrm{e}} 
\newcommand*\diff{\mathop{}\!\mathrm{d}}

\DeclareMathOperator{\sgn}{sgn}

\newcommand{\brackethyb}[1]{\{#1\}_{\vk}}

\newcommand{\normord}[1]{\xcentcolon\mathrel{#1}\xcentcolon}
\newcommand{\xcentcolon}{%
  \mathrel{\vbox{\hbox{$:$}\kern.2ex}}%
}

\newcommand{\todo}[1]{\textcolor[rgb]{1, 0.28, 0.1}{[ToDo: #1]}}
\newcommand{\MY}[1]{\textcolor[rgb]{1, 0.28, 0.1}{#1}}
\newcommand{\YW}[1]{\textcolor{blue}{#1}}

\begin{document}

\title {Coadjoint-orbit bosonization of a Fermi surface in a weak magnetic field}

\author{Mengxing Ye}
\email[email address: ]{mengxing.ye@utah.edu}
\affiliation{Department of Physics and Astronomy, University of Utah, Salt Lake City, UT, USA}

\author{Yuxuan Wang}
\email[email address: ]{yuxuan.wang@ufl.edu}
\affiliation{Department of Physics, University of Florida, Gainesville, FL, USA}
\begin{abstract}
We present a bosonized effective field theory for a 2d Fermi surface in a weak magnetic field using the coadjoint orbit approach, which was recently developed as a nonlinear bosonization method in phase space for Fermi liquids and non-Fermi liquids. We show that by parametrizing the phase space with the guiding center and the mechanical momentum, and by using techniques in noncommutative field theory, the physics of Landau levels and Landau level degeneracy ($N_{\Phi}$) naturally arises. For a parabolic dispersion, the resulting theory describes $N_{\Phi}$ flavors of free chiral bosons propagating in \emph{momentum space}. In addition, the action contains a linear term in the bosonic field, which upon mode expansion becomes a topological $\theta$-term. By properly quantizing this theory, we reproduce the well-known thermal and magnetic responses of a Fermi surface, including linear-in-$T$ specific heat, Landau diamagnetism, and the de Haas-van Alphen effect. {In particular, the de Haas-van Alphen effect is shown to be a direct consequence of the topological $\theta$-term.} Our theory paves the way toward understanding correlated gapless fermionic systems in a magnetic field using the powerful approach of bosonization.
\end{abstract}
\date{\today}
\maketitle

\tableofcontents

\section{Introduction}

One of the most remarkable features of gapless fermions at finite density is that low-energy excitations are localized near a momentum-space submanifold, i.e., the Fermi surface, which determines its physical properties and instabilities. One example is the Luttinger liquid~\cite{luttinger1963, haldane1981, vonDelft1998,Senechal1999,Khveshchenko1994} in one dimension (1d), in which the low-energy exications live near the Fermi points. From this insight,  the particle-hole excitations near the Fermi points were found to be described by a bosonic effective field theory (EFT), which becomes free in the low-energy limit. Such a bosonization procedure provides a powerful tool in understanding 1d gapless fermions, even in the strong coupling limit~\cite{luther-emery, haldane1981}.

In higher dimensions, a paradigmatic description of interacting gapless fermions is Landau's Fermi liquid theory~\cite{Landau1956}. Like its counterpart in 1d, the relevant degrees of freedom come from the vicinity of the Fermi surface (FS). It has been shown to great success that properties of the system only depend on a few microscopic parameters, e.g. the effective mass and the Landau parameters. While the Fermi liquid theory was not formulated as a quantum field theory, great efforts have been made to `bosonize' the Fermi liquid to recast it in a local EFT framework~\cite{HaldaneBosonization2005,FradkinBosonization1994,FradkinBosonizationPRL, khvesh,Houghton_2000,DDMS2022,kim2023,balents2024,khveshchenko2024,ravid2024electronslostphasespace}. For strongly correlated systems, e.g., Fermi surfaces coupled to critical bosons, Fermi liquid theory fails. A proper theoretical description of such non-Fermi liquids remains a major challenge, and typical strategies involve discretizing the Fermi surface into a number of patches~\cite{metlitski-sachdev-ising,sslee-review}, and extending the theory to large-$N$~\cite{AltshulerIoffeMillis1994, acs,metlitski-sachdev-ising, metlitski-sachdev-sdw, SYKreview} or introducing an extra small parameter $\epsilon$ {via dimensional or dynamical tuning}~\cite{nayak-wilczek,SenthilShankar2009,sslee-review}. Bosonization of patch fermions has also been applied to non-Fermi liquids~\cite{Houghton_2000}. However, it is generally difficult to accommodate Fermi surface curvature effects~\cite{khvesh} in such a ``patched" bosonization scheme (for Fermi or non-Fermi liquids). 
On the other hand, as the Fermi surface remains intact even for non-Fermi liquids, the nonlinear bosonization approach recently developed~\cite{DDMS2022} describing the dynamics of the entire Fermi surface in phase space may provide a promising alternative~\cite{DDMS2022,kim2023} towards a better-controlled theory of non-Fermi liquids.

Phenomenologically, compared with 1d Luttinger liquids, a powerful new probe into the gapless fermionic systems in higher dimensions is responses to magnetic fields. Orbital magnetic fields turns the gapless dispersion relation into dispersionless Landau levels (LLs) with spacing $\Delta E= \hbar \omega_c$, where $\omega_c$ is the cyclotron frequency. At strong fields, the LLs account for quantum Hall effects (with the help of disorder and possibly interaction effects). At weak fields, the LL spacing $\omega_c\ll \mu$, where $\mu$ is the Fermi energy, such that the notion of the Fermi surface {becomes fuzzy but} remains valid.
In this scenario the magnetic responses can be thought to capture the low-energy properties of the FS. Specifically, for $\omega_c\ll T$, the system displays Landau diamagnetism, a linear response to the orbital magnetic field $\bm B$ in the opposite direction, which for free fermions is proportional to the density of states at the FS. For $\omega_c\gg T$, the system displays oscillations in magnetic susceptibility that is periodic in $1/B$ with a period $\delta(1/B)=2\pi/\mathcal{A}_{\rm FS}$, where $\mathcal{A}_{\rm FS}$ is the size of the FS. This phenomenon is known as the de Haas-van Alphen (dHvA) effect or quantum oscillation, which is commonly used to experimentally probe the size and shape (in 3d) of the FS~\cite{Shoenberg1984}. 

For free fermions, the easiest approach to obtain the results above is simply summing over the LLs~\cite{LK1956}.
For interacting fermions, while a theoretical treatment is more challenging (see e.g., Refs.~\cite{Nosov2024,Raines-Chubukov,Else2021PRX}), these responses are expected to persist. For example, in Ref.~\cite{Else2021PRX}, from the perspective of emergent symmetries and UV-IR anomaly matching, it has been argued that the period of dHvA oscillations remain the same, proportional to $1/\mathcal{A}_{\rm FS}$. The dHvA oscillation can also be understood semiclassically via Bohr-Sommerfeld quantization~\cite{Luttinger1951,Onsager1952,LK1956}. The phase accumulated by a quasiparticle as it moves in a semiclassical orbit {in momentum space around the FS}, which depends on $\mathcal{A}_{\rm FS}\ell_B^2$ plus corrections~\cite{Kohn1959,Roth1962,Blount1962} due to e.g. the Berry curvature~\cite{Mikitik1999,Alexandradinata2023ar} enclosed inside a FS, is quantized, which leads to 
{the period of dHvA oscillations.}



Motivated by previous results showing that the dHvA effect can be captured by low-energy parameters near the FS~\cite{Luttinger1961,Gorkov-1962,Nosov2024,Else2021PRX}, it is desirable to develop a fully-fledged EFT approach to describe the dHvA effect,
which can reveal the connections between LL quantization, the phase of a semiclassical orbit, and the emergent FS anomaly. Moreover, the effects of interactions can be studied systematically within this field theory framework.  Bosonization of the FS~\cite{HaldaneBosonization2005,FradkinBosonization1994,FradkinBosonizationPRL, khvesh,Houghton_2000,DDMS2022,kim2023,balents2024} is an obvious candidate for such an EFT. 

However, to the best of knowledge,  thirty years after its conception, bosonization approaches have not yet been rigorously tested in the context of magnetic responses. Notably, in Ref.~\cite{Fradkin2018}, the authors bosonized a Fermi liquid in a weak $B$ field, and were able to obtain a magnetic response with oscillatory behavior in $1/B$. However, the period of the oscillation in $1/B$ obtained there is inversely proportional to an undetermined UV cutoff $\Lambda\ll k_F$ (thus displaying  a pathological UV-IR mixing), which is much larger than $2\pi/\mathcal{A}_{\rm FS}$.  A closer look from the perspective of field theory  reveals that the effect of magnetic field are highly nontrivial. First, in field theories, the magnetic field couples to the system via the vector potential $\bm A$, rather than the field strength $\bm B = \nabla\times\bm A$. 
In an arbitrarily small uniform magnetic field, it results in Landau level wave functions that cannot be perturbatively connected to plane waves. It implies that even in the ``linear-response" (Landau diamagnetism) regime $\omega_c\gg T$, a perturbative treatment via the standard linear response theory can be challenging~\cite{RMPNenciu1991}\footnote{One may consider instead a staggered magnetic field and then take the $\vq\to 0$ limit to obtain the diamagnetic response; see  Ref.~\cite{Mayrhofer2024} and references therein.}.
Furthermore, not only is dHvA effect nonlinear in $B$, the response function $\sim \cos(\mathcal{A}_{\rm FS}/B)$ features an essential singularity at $B=0$, which cannot be obtained by a response theory to any order in $B$.

\subsection{Methods and Summary}
In this paper, we develop a systematic procedure of bosonization of FS's in the presence of a weak magnetic field $\omega_c\ll \mu$. (Note that this is in a opposite regime of bosonization of the lowest LLs.~\cite{Conti1998,Karabali2004, Karabali2006,Girvin2005,Son2016}) We will focus on two spatial dimensions (2d) with a magnetic field in the perpendicular direction. We expect it to be  straightforward to generalize our results to 3d, where the electron motion in the third direction is unaffected by the magnetic field.

We do so by following the nonlinear bosonization via the method of coadjoint orbits recently introduced in Refs.~\cite{DDMS2022,MehtaThesis2023} (see also \cite{Khveshchenko1994}), which gives the effective action in terms of the fluctuations (denoted as $\phi$) of the shape of a Fermi surface from its ground state phase space configuration. An unusual feature of the bosonized action is that the base manifold is the \emph{phase space} $(\bm x, \bm p)$, while the target manifold is the \emph{coadjoint orbits} formed by deformations of the ground state single-particle distribution function $f_0(\bm x, \bm p)$ as Lie group elements. The corresponding Lie algebra is the Moyal algebra; in the semiclassical limit it reduces to the Poisson algebra, and the Lie group becomes the group of familiar canonical transformations. Compared with earlier works~\cite{FradkinBosonization1994,FradkinBosonizationPRL, HaldaneBosonization2005,Houghton_2000} on higher-dimensional bosonization, a key feature of this method is that it gives rise to an action that is inherently nonlinear. Upon expansion in the boson fields, it becomes an EFT with infinitely many irrelevant couplings.

A weak magnetic field can be incorporated to the bosonized theory via minimal coupling, and one straightforward way to proceed is to treat it as a perturbation~\cite{Huang2024}. However, as we argued above, no perturbative expansion can reproduce the dHvA effect with an essential singularity at $B=0$. Instead, we show that magnetic field modifies the bosonized action at a more fundamental level. In particular, to obtain the correct integer-valued LL degeneracy $N_{\phi}$, we demonstrate that with an arbitrarily small magnetic field, the bosonized action needs to be taken beyond its semiclassical limit, such that even in the ``semiclassical" regime $T,\omega_c\ll \mu$, one needs to begin with the full Moyal algebra rather than replacing it with the Poisson algebra. In addition, on a torus the magnetic flux is quantized, and the magnetic translation symmetry puts constraints to the theory. To take it into account, it is convenient to work with phase space basis consisting of the guiding center $\bm R$, which generates magnetic translations, and the mechanical/physical momentum $\bm k$. In these new coordinates, the Moyal algebra of $\vk$ can be truncated as Poisson algebra in $k_x$ and $k_y$ (the two perpendicular components to the magnetic field) in the semiclassical regime. The coordinates $R_x$ and $R_y$, on the other hand, are still on Moyal algebra and need to be treated more carefully, which we do by mapping the theory to a noncommutative field theory~\cite{Douglas2001rmp,Szabo2003} and keeping the relevant modes. {Such a treatment naturally leads to $N_\Phi$ independent boson modes each subject to the same action, reproducing the LL degeneracy.}

As a key result of this procedure, we obtain an action of $N_\Phi$ flavors of 1d chiral bosons $\phi_i(\theta_{\vk}, t)$~\cite{Fradkin2018,Caldeira1997} that lives in \emph{momentum space}, where $\theta_{\vk}$ is an angle parametrizing the FS.  Furthermore, when the dispersion is parabolic, the chiral bosons become non-interacting, in sharp constrast with the case without a magnetic field~\cite{DDMS2022}. In other words, in a magnetic field, the cyclotron motion around the FS can be thought of as quantum Hall edge states in momentum space. The corresponding chiral anomaly can be captured by a Chern-Simons response theory in momentum space. This is directly related to the recent proposal~\cite{Else2021PRX} of a LU(1) anomaly associated to a FS, which, among other things, led to an elegant nonperturbative proof of the Luttinger theorem. 

{An important insight of this work is that, to properly quantize the chiral boson action on compact space $\theta_{\vk}$, one needs to incorporate the winding modes of the compact fields $\phi_i(\theta_{\vk}, t)$.} The Hilbert space is then divided into sectors with different winding numbers that differ by integers. These integers correspond to the numbers of additional electrons/holes from the system.  Within each sector, the Hilbert subspace is a Fock space for bosons with energy $\omega_c$. At the quantum level the bosonized action is capable of describing the system in a grand canonical ensemble, which we directly prove by computing the partition function. In 1d bosonization, adding and removing electrons is explicitly acheived by the vertex operators (with the necessary Klein factors)~\cite{vonDelft1998,Senechal1999}.  For a FS it remains to be seen whether and how vertex operators can be constructed. 

As a main difference with earlier literature on FS bosonization~\cite{HaldaneBosonization2005,FradkinBosonization1994, DDMS2022, Fradkin2018} and bosonization theories starting from LLs~\cite{Caldeira1997}, we uncover an additional term {from the Weiss-Zumino-Witten term of} the action that is linear in $\phi$, which was previously argued to be a total derivative and dropped. However, we show that this term cannot be neglected due to the winding modes. When reduced to $(0+1)$-d (since the $\theta_{\vk}$ direction is compact and the corresponding modes are discrete), this term is a topological $\theta$-term. Indeed, with this $\theta$-term the action resembles that of a particle moving on a ring subject to a fictitious flux $\sim \mathcal{A}_{\rm FS}/B$ that is inverse of the physical magnetic flux, and the ground-state energy shows oscillatory behavior that is $2\pi$-periodic in $\sim \mathcal{A}_{\rm FS}/B$. At $T\lesssim \omega_c$, this gives precisely the dHvA effect, which remarkably is a topological effect in the bosonized language and indeed elusive in any perturbative expansion in $B$. 

At $T\gg \omega_c$, the thermal energy can be expressed as a sum of average ground state energy of different winding sectors and the
``black-body radiation" from the bosons. From this, one obtains the specific heat for a Fermi surface $C=\pi^2g(E_F)T/3$, where $g(E_F)$ is the density of states at the Fermi surface. We note that in the bosonization scheme without magnetic field, the specific heat has only been obtained to be linear in $T$, while the precise prefactor involves UV-IR mixing and remains undecided.~\cite{DDMS2022} In this context, the magnetic field in our theory acts as a UV regulator, enabling the precise evaluation of specific heat (likely well beyond the free-fermion limit).
For magnetic response, we show that in this regime the oscillatory behavior is exponentially suppressed, consistent with the Lifshitz-Kosevich formula, while the regular dependence of $B$ gives rise to Landau diamagnetism. We note that while our analysis on magnetic responses in this paper focuses on free fermions with parabolic dispersion, our bosonization scheme can be extended to a generic band with nonzero Berry curvature~\cite{Mikitik1999,Alexandradinata2023ar}, Fermi liquids and non-Fermi liquids~\cite{DDMS2022,khveshchenko2024}, which we leave to future work.

{The remainder of this paper is organized as follows.} In Sec.~\ref{sec:moyal} we review the coadjoint-orbit approach to nonlinear bosonization of a FS with Moyal algebra. In Sec.~\ref{sec:mag} we analyze the bosonized action in the presece of weak magnetic field. In Sec.~\ref{sec:quant} we quantize the action and show that the familiar physics of degenerate LLs naturally emerges from the bosonized theory. In Sec.~\ref{sec:resp}, we directly obtain the thermal and magnetic response functions of a 2d FS from the bosonized theory. 

\section{Construction of the Effective Action}
\label{sec:moyal}

In this work we follow Refs.~\cite{DDMS2022,MehtaThesis2023} and consider the bosonization of Fermi surfaces in 2d using the method of coadjoint orbits.
In Ref.~\cite{DDMS2022}, the degrees of freedom considered are the canonical transforms of a Fermi surface, i.e., the volume-preserving deformations thereof. Formally, these canonical transforms form a Lie group, with the Lie algebra  being the Poisson algebra. 
To obtain the effective action for a Fermi surface in a weak magnetic field, we argue that one should instead focus on the Moyal algebra $\mathfrak{g}$ and the associated Lie group $\mathcal{G}$, which reduces to Poisson algebra only in the limit $\hbar \overleftarrow{\nabla}_{\ve{x}}\overrightarrow{\nabla_{\ve{p}}}\ll 1$ (in all other equations we have set $\hbar=1$). However, we found that this does not apply even in the ``semiclassical" regime $\omega_c\ll \mu$. In particular,  to describe the LL degeneracy, Moyal algebra should be kept. 

{For this reason, we will derive the bosonized action on Moyal algebra.} We first review the basics of Moyal algebra and its associated coadjoint orbits. Next we present the bosonized action expressed in Moyal brackets. We then discuss its coupling to background gauge fields. 

We note that the discussion in this Section has appeared in Refs.~\cite{DDMS2022,MehtaThesis2023} using Poisson algebra, and many aspects involving Moyal algebra in FS bosonization have been addressed in Ref.~\cite{MehtaThesis2023,Umang_talk,Luca_talk}. In order to be self-contained, we present them in full details here and in Appendix \ref{app:path}.

\subsection{Moyal algebra and the coadjoint orbit}

We begin with a lightning review of the mathematical structures.
The Moyal algebra $\mathfrak{g}$ is defined via the Moyal bracket: For $F(\vx, \vp)$ and $G(\vx, \vp)$
\begin{align}
\{F, G\}_{\rm M} &= - \iu (F \star G - G \star F)\non\\
 &= 2 F \sin\left( \frac{\overleftarrow{\nabla}_\ve{x} \overrightarrow{\nabla_{\ve{p}}}-\overleftarrow{\nabla}_\ve{p} \overrightarrow{\nabla_{\ve{x}}}}{2} \right) G ,
 \label{eq:1}
\end{align}
with a star product defined as
\begin{equation}
    F \star G = F \exp\left( \iu \frac{\overleftarrow{\nabla}_\ve{x} \overrightarrow{\nabla_{\ve{p}}}-\overleftarrow{\nabla}_\ve{p} \overrightarrow{\nabla_{\ve{x}}}}{2} \right) G .
     \label{eq:2}
\end{equation} 
{It can be verified that the star product is associative.}
As shown in the next Subsection and in  Appendix \ref{app:path}, the star product and Moyal bracket appears naturally in Wigner representation of operator products~\cite{KamenevBook} and in the derivation of the bosonized action via coherent-state path integral.
We note that in the semiclassical limit $ \overleftarrow{\nabla}_{\ve{x}}\overrightarrow{\nabla_{\ve{p}}}\ll 1$, the Moyal bracket reduces to the Poisson bracket $\{F, G\}$.

As usual,  $\mathfrak{g}$ as a linear space form a representation of $\mathfrak{g}$ itself, i.e., the adjoint representation/action. For $G\in\mathfrak{g}$, the adjoint action on $\mathfrak{g}$ is given by
\begin{align}
{\rm ad}_G &: \mathfrak{g} \rightarrow \mathfrak{g}\non\\
{\rm ad}_G F &\equiv \{G, F\}_{\rm M},
\end{align}
such that
\be
{\rm ad}_H {\rm ad}_G - {\rm ad}_G {\rm ad}_H = {\rm ad}_{\{H, G\}_{\rm M}}.
\ee
which follows from the Jacobi identity.

In physical terms, elements in $\mathfrak{g}$ are (one-particle) observables $F( {\ve x},  {\ve p})$, and the expectation value is given by the inner product
\be
\Langle f,F\Rangle \equiv \int\frac{\diff \vx \diff \vp}{(2\pi )^2} f(\ve x, \ve p) F(\ve x, \ve p),
\label{eq:inner}
\ee
where $f(\ve x, \ve p)$ is the one-particle distribution function. The inner product $\Langle f,F\Rangle$ defines a dual linear space $\mathfrak{g}^*$, which is the linear space for distribution functions. Demanding $\Langle f,\adj{G} F \Rangle=-\Langle\coadj{G}{f},F\Rangle$
{for analytic functions $f$ and $F$}, one can similarly define a coadjoint action on $\mathfrak{g}^*$ is, for $G\in \mathfrak{g}$
\begin{align}
{\rm ad}^*_G &: \mathfrak{g}^* \rightarrow \mathfrak{g}^*\non\\
{\rm ad}^*_G f &\equiv \{G, f\}_{\rm M}.
\end{align}

The coadjoint action of the Lie algebra in turn generates the coadjoint representation of the Lie group. {For $U=\exp(-iG)$, the coadjoint action is}
\begin{align}
{\rm Ad}^*_{U} f & \equiv f + {\rm ad}^*_G f + \frac{1}{2!} {\rm ad}^*_G \({\rm ad}^*_G f\) + \cdots \non \\
& = f + \{G, f\}_{\rm M} + \frac{1}{2!}  \{ G, \{G, f\}_{\rm M} \}_{\rm M} + \cdots \non \\
& = U \star  f \star U^{-1},
\label{eq:coAdj}
\end{align}
{satisfying $\Langle f, F \Rangle = \Langle \coAdj{U} f, \Adj{U} F \Rangle $.}

Denoting the ground state distribution function as $f_0(\ve x, \ve p)= \Theta (p_F - \abs{\vp})$, where $p_F$ is the Fermi wavevector, its \emph{coadjoint orbit} $\mathcal{O}_{f_0}$ is defined as the set of all elements $f\in \mathfrak{g}^*$ that can be obtained by the coadjoint action on $f_0$, $\mathcal{O}_{f_0}$,
\begin{align}
\mathcal{O}_{f_0}=\{f={\rm Ad}^*_{U} f_0 | U\in \mathcal{G}\}.
\label{eq:f0B0}
\end{align}
$f_0$ is invariant under a subgroup of canonical transformations, denoted as $\mathcal{H}$, where 
\begin{equation}
\mathcal{H}= \{ V \in \mathcal{G} | {\rm Ad}^*_V f_0 = f_0\}
\label{eq:stablizer}
\end{equation}
so ${\rm Ad}^*_{U\star V} f_0 = {\rm Ad}^*_{U} f_0$. There is a one-to-one mapping between $f$ in the coadjoint orbit $\mathcal{O}_{f_0}$ and the left coset $U \mathcal{H}$, so we have $\mathcal{O}_{f_0} \cong \mathcal{G}/\mathcal{H}$.

The coadjoint orbit $\mathcal{O}_{f_0}$ constitutes the target manifold of our bosonized EFT, which we discuss below.

\subsection{Coadjoint-orbit action with Moyal algebra}
\label{sec:IIB}
In Ref.~\cite{DDMS2022}, the authors constructed an effective action on the coadjoint orbit (with Poisson instead of Moyal algebra), which leads to, as the equation of motion, the collisionless Boltzmann equation. 

Similarly, here the action of coajoint orbits $\mathcal{O}_{f_0}$ is given by two parts. First, the dynamic phases of the coadjoint orbits is generated by a Hamiltonian,
\begin{align}
H[f]=&\Langle f, \epsilon(\vp) \Rangle + H_{\rm LP}[f] \\
=&\Langle U(\vx, \vp)\star f_0\star  U^{-1}(\vx, \vp), \epsilon(\vp)  \Rangle + H_{\rm LP}[f],\non
\end{align}
where 
\be
f(\vx, \vp)=U(\vx, \vp) \star f_0(\vp) \star U^{-1}(\vx, \vp),
\label{eq:fMoyal}
\ee
is one-particle distribution function,
$\epsilon(\vp) = \epsilon_0(\vp) - \mu$ is the dispersion, and $H_{\rm LP}[f]$, which is of higher order in $f$,  captures interaction effects in terms of generalized Landau parameters.
Second, the coadjoint orbit is known to be a symplectic manifold with an exact and non-degenerate symplectic 2-form, known as the Kirillov-Kostant-Sourian (KKS) 2-form~\cite{DDMS2022}. Just like the path integral of spin~\cite{altland}, the KKS 2-form can be used to construct a Weiss-Zumino-Witten (WZW) term. This WZW term captures the Berry phases of coadjoint orbits evolving in time. Combined, the action is given by~\cite{DDMS2022}
\begin{align}
S[f] &= S_{\rm WZW}[f] + S_H[f] , \non\\
S_{\rm WZW}[f] &= \int \diff t \Langle f_0, U^{-1} \star \iu \partial_t U \Rangle,\non\\
S_{H}[f] &= -\int \diff t \Langle f, \epsilon(\vp) \Rangle + S_{\rm LP}[f].
\label{eq:11}
\end{align} 
where $S_{\rm LP}[f] = -\int dt H_{\rm LP}[f]$ incorporates interaction effects. {For the remainder of this work, we will focus on non-interacting fermions with $S_{\rm LP}=0$, while postpone the analysis of Landau parameters in a Fermi liquid to a future work.}

While $S[f]$ in general should be viewed as an EFT, which captures the renormalization group flow of the microscopic theory in the IR, we show that for \emph{free} fermions, this action, in particular the WZW term, can be explicitly derived from the microscopic theory path integral formalism using the coherent state of the Fermi surface~\cite{FradkinBosonization1994,Luca_talk,Umang_talk}, in which $U(\vx, \vp)$ is a Wigner function and the Moyal algebra naturally emerges. We show this derivation in Appendix \ref{app:path}.

We also note that the equation of motion $\delta S[f]=0$ for noninteracting particles is given by
\be
\partial_t f(\vx,\vp) = \coadj{\epsilon(\vp)} f(\vx, \vp),
\ee
which is precisely the collisionless \emph{quantum Boltzmann equation}~\cite{KamenevBook}. While it reduces to the classical Boltzmann equation in the semiclassical limit, as we mentioned, we will remain in the quantum regime, which is crucial for obtaining the LL degeneracy.

\subsection{Coupling to the background gauge field}
The global $U(1)$ symmetry associated with the charge conservation acts through $\lambda(\vx,\vp,t)=\lambda(t) \in \mathfrak{g}$, where $\lambda(t)$ is homogeneous in space. To gauge it, the theory should be invariant under local $U(1)$ transformation, so $\lambda(\vx,\vp, t)=\lambda(\vx, t) \in \mathfrak{g}$ is space-time dependent. The gauge invariant action reads~\cite{DDMS2022}
\begin{align}
S &= S_{\rm WZW} + S_{H}, \non\\
S_{\rm WZW} &= \int \diff t \Langle f_0, U^{-1} \star \[i \partial_t - e A_0 \star\] U \Rangle,\non\\
S_{H} &= -\int \diff t \int \frac{\diff \ve{x}\diff\ve{p}}{(2\pi)^2}f(\vx, \vp, t) \epsilon(\vp+e \ve{A}).
\label{eq:actiongauged}
 \end{align}   
 The gauge transformation is
\begin{equation}
U \rightarrow W \star U ,\quad A_\mu \rightarrow W^{-1} \star (A_\mu + \frac{i}{e} \partial_\mu) \star W
\label{eq:GaugeTransform}
\end{equation}
with $W = \exp (i \lambda(\vx, t))$. We have used $\coAdj{W^{-1}} f(\vx, \vp, t) = f(\vx, \vp - \nabla \lambda, t)$ for the function $f$ on Moyal algebra. Eq.~\eqref{eq:actiongauged} resembles the gauged effective action in Ref.~\cite{DDMS2022}, but replace the multiplication and Poisson bracket with Moyal product and bracket. 



\section{Bosonic action in a weak magnetic field}
\label{sec:mag}

In this section, we focus on the theory coupled to a background magnetic field , and use the symmetric gauge, i.e. $A_0 = 0$ and $\ve{A} = - \frac{B}{2} \vx \times \hat{\ve{z}}$.  
{To study the dynamics of the excitations near the Fermi surface through Eq.~\eqref{eq:actiongauged}, we first rewrite the action in terms of a new set of phase space coordinates $(\ve{R},\ve{k})$, where $\vR$ is the guiding center coordinate, $\vk$ is the physical momentum. The action can be simplified in the limit $\mu\gg \omega_c$ and by imposing the magnetic translation symmetry constraints, which are discussed in Secs.~\ref{sec:hyb} and~\ref{sec:MagneticTranslation}, respectively. In Sec.~\ref{sec:ChiralBosonAction}, we present the effective action in $\vk$-space, and discuss its connection with the phase space Chern-Simons action that describes the LU(1)-anomaly for Fermi surfaces in Ref.~\cite{Else2021PRX}.}

\subsection{The \texorpdfstring{$\(\bm{R}, {\bm k}\)$}{(R,k)} coordinates of phase space}
\label{sec:hyb}

To proceed, we first introduce a set of new variables $({\vR}, \ve{k})$ through 
\begin{align}
    \ve{R}=\ell_B^2 (\ve{p}-e\ve{A})\times \hat{\ve{z}},\quad
    \ve{k}=\ve{p}+e\ve{A}
\end{align}
where $\ell_B=1/\sqrt{B}$ is the magnetic length, $\ve{R}$ is a constant of motion and corresponds to the guiding center coordinate in the symmetric gauge, and $\ve{k}$ corresponds to the mechanical momentum. With $\ve{A}=-B \ve{x}\times \hat{\ve{z}}/2$,
$({\vR}, \ve{k})$ can be viewed as phase space coordinates in a new basis. Compared with earlier works, in these coordinates the magnetic translation symmetry is transparent, since $\ve{R}$ is proportional to the generator of magnetic translation symmetry.

The $({\bm{R}}, {\bm k})$ coordinates satisify the Moyal brackets
\be
\{{\ve{R}}_i, {\ve{R}}_j\}_{\rm M} = \epsilon_{ij} \ell_B^2, \{ \vk_i, \vk_j\}_{\rm M} =-\epsilon_{ij} \ell_B^{-2}, \{ \vR_i, \vk_j\}_{\rm M} =0.
\ee
and it is straightforward to check that the star product in $({\bm{R}}, {\bm k})$ reads
 \begin{align}
  F \star G &= F \exp \left( \iu\frac{\overleftarrow{\partial}_\ve{R}\times\overrightarrow{\partial_{\ve{R}}}}{2B} \right) \otimes \exp\left( -\iu B \frac{ \overleftarrow{\partial}_\ve{k} \times \overrightarrow{\partial_{\ve{k}}}}{2} \right) G  \non\\
  &\equiv F \star_{\vR}\star_{\vk} G.
 \end{align}    
Notably there are no cross terms such as $\overleftarrow{\partial}_{\ve{k}} \overrightarrow{\partial}_{\ve{R}}$ in the Moyal product, and $\star_{\vk}$ and $\star_{\vR}$ commute. {Define all functions $\mathcal{F}(\vx, \vp,t)$ in the $(\vR, \vk)$ coordinates as $\mathcal{F}'(\vR, \vk,t)=\mathcal{F}(\vx, \vp, t)$,} the inner product \eqref{eq:inner} can be rewritten as
\be
\Langle f,F\Rangle= \int\frac{\diff \vk \diff \vR}{(2\pi)^2} f'(\vR, \vk) F'(\vR, \vk).
\ee
For brevity, the primes will be dropped hereafter. The effective action Eq.~\eqref{eq:actiongauged} for the Fermi surface in the homogeneous magnetic field in the $(\vR, \vk)$ coordinates becomes
\begin{align}
S = \int \diff t \,\Langle f_0, \iu U^{-1} \star_\vR \star_\vk \partial_t  U \Rangle -\Langle f, \epsilon\Rangle
\label{eq:actiongauged1}
 \end{align}   
where $\epsilon(\vR, \vk)=\epsilon(\vk)$, and $f_0$ is the equilibrium distribution given by the Hamiltonian $\epsilon(\vk)$.

To proceed, while it may be tempting to expand the exponential operator in the star product $\star_{\vR}$, this approach is not justified, even in the semiclassical regime, {as typically the argument of the exponential in $\star_\vR$ is $B^{-1}\overleftarrow{\partial}_{\ve{R}}\times \overrightarrow{\partial}_{\ve{R}} \sim 1$}. 
Instead, the $\star_{\vR}$ product in the bosonized action  can be treated using standard methods of non-commutative field theories with a base manifold where the $\vR$ coordinates are replaced by operators $\widehat{\ve{R}}$ obeying the commutation relation $\left[ \widehat{\ve{R}}_i, \widehat{\ve{R}}_j\right] =\iu \epsilon_{ij} \ell_B^2$. 

To this end we make use of the identity
\begin{align}
&\int {\diff \ve{R}} \mathcal{F}_1(\ve{R}) \star_\vR \mathcal{F}_2 (\ve{R}) \star_\vR ... \star_\vR \mathcal{F}_n (\ve{R}) \non\\
=& 2 \pi \ell_B^2 \Tr \left[ \mathcal{F}_1^{\rm W}(\widehat \vR) \mathcal{F}_2^{\rm W}(\widehat \vR) ... \mathcal{F}_n^{\rm W}(\widehat \vR) \right] 
\label{eq:NCommutativeAction}
\end{align}
where the trace $\Tr$ should be understood as integration over the noncommutative coordinates $\widehat\vR$, {which will be discussed in detail for a torus in Sec.~\ref{sec:MagneticTranslation}}. $\mathcal{F}^{\rm W}(\widehat \vR)$ is the Weyl transform~\cite{Szabo2003,Douglas2001rmp} of $\mathcal{F}(\ve{R})$ defined via
\begin{align}
\mathcal{F}^{\rm W}(\widehat \vR) &= \int \frac{\diff \vq}{2\pi \ell_{B}^{-2}} \tilde{\mathcal{F}}(\vq) \eu^{\iu \vq \cdot \widehat{\ve{R}}},\non\\
\tilde{\mathcal{F}}(\vq) &= \int \frac{\diff \vR}{2\pi \ell_B^2} \eu^{-\iu \vq \cdot \ve{R}} \mathcal{F}(\ve{R}).
\label{eq:21}
\end{align}
$\mathcal{F}^{\rm W}(\widehat \vR)$ can be viewed as the ``quantum version" of $\mathcal{F}(\ve R)$, and for all purposes can be simply denoted as $\mathcal{F}_{\widehat\vR}$. Inverting the Weyl transform, $\mathcal{F}(\vR)$ is the Wigner function of $\mathcal{F}_{\widehat \vR}$, and from this perspective Eq.~\eqref{eq:NCommutativeAction} follows directly from the property of Wigner functions (cf.\ Eq.~\eqref{eq:a8}).
Further using the property $\int_{\vR} F G = \int_{\vR} F \star_\vR G$, we can re-express the action \eqref{eq:actiongauged1} as
\be
S = \int \diff t\Tr\[\Langle f_{0,\widehat \vR}, U_{\widehat\vR}^{-1}\star_{\vk}\iu \partial_t U_{\widehat\vR} \Rangle_{\vk}-\Langle f_{\widehat \vR},\epsilon(\vk)\Rangle_{\vk}\],
\label{eq:23}
\ee
where $f_{\widehat\vR}=U_{\widehat\vR}\star_{\vk} f_{0,\widehat\vR}\star_{\vk}U^{-1}_{\widehat\vR}$, and we have defined the inner product 
\be
\Langle f, F\Rangle_{\vk}=\int\frac{\diff \vk}{2\pi B} f(\vk) F(\vk).
\label{eq:innerk}
\ee

In the semiclasscal regime, we note that {for typical low-energy fluctuations}
\begin{align}
B \overleftarrow{\partial}_{\ve{k}} \times \overrightarrow{\partial}_{\ve{k}} \sim \frac{B}{k_F^2} \sim \frac{\omega_c}{\mu}\ll 1,
\label{eq:25}
\end{align} and one can expand and truncate the $\star_\vk$ product, and the terms in Eq.~\eqref{eq:23} can be approximated with sum of nested Poisson brackets. Parametrizing $U$ as 
\be
U_{\widehat\vR}(\vk, t) = \exp[\iu \phi_{\widehat\vR} (\vk, t)],
\ee
the distribution function $f_{\widehat\vR}=U_{\widehat\vR}\star_{\vk} f_{0,\widehat\vR}\star_{\vk}U^{-1}_{\widehat\vR}$ after the truncation reads, 
\begin{align}
f _{\widehat\vR}(\vk, t) \approx& f_{0,\widehat\vR}(\vk) - \{\phi_{\widehat\vR}, f_{0,\widehat\vR}\}_{\vk} \non\\
&+\frac{1}{2} \{\phi_{\widehat\vR}, \{\phi_{\widehat\vR}, f_{0,\widehat\vR}\}_{\vk}\}_{\vk} + \mathcal{O}(\phi^3),
\label{eq:fHybrid}
\end{align}
The product $U^{-1}_{\widehat\vR}(\vk, t)   \star_{\vk} \partial_t U_{\widehat\vR}(\vk, t)$ reads~\cite{stackexchange}
\begin{align}
U^{-1}_{\widehat\vR}(\vk, t)   \star_{\vk} &\partial_t U_{\widehat\vR}(\vk, t) \approx-\dot{\phi}_{\widehat\vR} - \frac{1}{2} \brackethyb{\phi_{\widehat\vR}, \dot{\phi}_{\widehat\vR}} \non\\
& - \frac{1}{6} \brackethyb{\phi_{\widehat\vR}, \brackethyb{\phi_{\widehat\vR}, \dot{\phi}_{\widehat\vR}}} + \mathcal{O}(\phi^4),
\end{align}
where the Possoin bracket $\{F,G\}_{\vk}$ is defined as
\be
\{F,G\}_{\vk} = -B\, \partial_{\vk} F \times \partial_{\vk} G.
\label{eq:poissonk}
\ee
 In the remainder of this work, we will omit the subscript for $\{,\}_{\vk}$ and $\langle , \rangle_{\vk}$ unless otherwise specified.

\subsection{Evaluating  the trace over noncommutative coordinates\label{sec:MagneticTranslation}}
We now proceed to evaluate the trace in Eq.~\eqref{eq:23} over the noncommutative coordinates $\widehat \vR$. A convenient basis turns out to be the eigenbasis of magnetic translation symmetry. Importantly, magnetic translation symmetry also imposes strong constraints on the target manifold of the theory, i.e.,  coadjoint orbit $\mathcal{O}_{f_0}$, which further simplifies the evaluation of the trace.

\subsubsection{Magnetic translation symmetry}
The magnetic translation operator is defined as 
\begin{align}
\widehat t(\ve{d})=\exp[-\iu \frac{d_x \widehat R_y - d_y \widehat R_x}{\ell_B^2}] 
\end{align}
where $\ve{d} \in \mathbb{R}^2$. The product of $\widehat t(\ve{d})$ satisfies  $\widehat t(\ve{d}_1)\widehat t(\ve{d}_2) = \exp(\frac{\iu}{2} (\ve{d}_1 \times \ve{d}_2)_z /\ell_B^2)\widehat t(\ve{d}_1+\ve{d}_2) $, where $(\ve{d}_1 \times \ve{d}_2)_z=(\ve{d}_1 \times \ve{d}_2)\cdot \hat{\ve{z}}$.

In the presence of a magnetic field, even though all $\widehat t({\ve{d}})$'s commute with the Hamiltonian, they generally do not commute with themselves except for special values of $\ve d$'s.  
As a result, for an electron gas with continuous translation symmetry, the magnetic field breaks it to commuting and discrete magnetic translation symmetry, with sizes  $\ve{\ell}_1$ and $\ve{\ell}_2$ such that 
\begin{align}
\widehat t(\ve{\ell}_1)\widehat t(\ve{\ell}_2) = \exp[\iu (\ve{\ell}_1 \times \ve{\ell}_2)_z /\ell_B^2]\widehat t(\ve{\ell}_2)\widehat t(\ve{\ell}_1)=\widehat t(\ve{\ell}_2)\widehat t(\ve{\ell}_1).
\end{align} 
The condition is
\be
|\ve{\ell}_1 \times \ve{\ell}_2| = 2\pi \ell_B^2.
\ee
The vectors $\ve{\ell}_{1,2}$ define a \emph{magnetic unit cell}, which  encloses a magnetic flux quanta. Without loss of generality, we assume $\ve{\ell}_1 \perp \ve{\ell}_2$, and a system of length $L_1=M_1 |\ve{\ell}_1|$ and width $L_2=M_2 |\ve{\ell}_2|$ on a torus. 
Thus, the set of all commuting magnetic translation operators is given by 
$\mathbb{T}=\{\widehat t(\ve{\ell})| \ve{\ell}= n_1 \ve{\ell}_1 + n_2 \ve{\ell}_2; n_i = 1, 2, ..., M_i\}$. 
From $\mathbb{T}$, we can obtain a set of orthogonal and complete basis functions that simultaneously diagonalizes \emph{all} magnetic translation operators in $\mathbb{T}$. Following Ref.~\cite{Haldane2018b}, it can be constructed by first identifying the zero momentum state satisfying 
\begin{align}
\widehat t(\ve{\ell}) \ket{\ve{0}} = \xi(\ve{\ell}) \ket{\ve{0}}
\end{align} 
where $\xi(\ve{\ell})=1$ if $\widehat t(\ve{\ell}/2)\in \mathbb{T}$, $\xi(\ve{\ell})=-1$ otherwise. The finite momentum eigenstates are generated by 
\begin{align}
    \ket{\ve{K}}= \eu^{\iu \ve{K} \cdot \widehat{\ve{R}}} &\ket{\ve{0}},\non\\
    \ve{K} = \frac{2\pi n_1 }{ L_1} \hat{\ve{x}} + \frac{2\pi n_2}{ L_2} \hat{\ve{y}},&~~ n_i = 1, 2, ..., M_i,
\end{align}
i.e.\ the $\ve{K}$'s are crystal momenta in the first magnetic Brillouin zone (mBZ), whose total number is 
\be
N_\Phi = M_1M_2= \frac{BL_1L_2}{2\pi},
\ee
which is the total flux quanta through the entire system.

This definition ensures that 
\be
\widehat t(\ve{\ell}) \ket{\ve{K}} = \xi(\ve{\ell}) \eu^{\iu \ve{K} \cdot \ve{\ell}}\ket{\ve{K}}
\label{eq:35}
\ee
and $\braket{\ve{K}'} {\ve{K}} = \delta_{\ve{K}'\ve{K}}$, and thus $\{\ket{\Psi(\ve{K})}| \ve{K} \in \text{mBZ}\}$ defines a complete orthogonal basis in the non-commutative manifold $\widehat{\ve{R}}$. Therefore, the trace in Eq.~\eqref{eq:NCommutativeAction} can be expressed as 
\begin{align}
\Tr[...] = \sum_{\ve{K} \in {\rm mBZ}} \bra{\ve{K}} ... \ket{\ve{K}}.
\end{align}

\subsubsection{Constraining the coadjoint orbit with the magnetic translation symmetry}
\label{sec:constraint}

The coadjoint orbit $\mathcal{O}_{f_0}$, {describes all deformation of the one-particle distribution function $f_{0,\widehat\vR}(\vk,t)$ pertinent to the Hamiltonian evolution. Since the Hamiltonian are symmetric under magnetic translation, we can  further constrain (see Appendix \ref{app:path} for details)}
$\mathcal{O}_{f_0}$ {by requiring distrubution functions $f_{\widehat{\vR}}(\vk,t)$ in the target manifold} to be invariant under all  magnetic translation operators in $\mathbb{T}$ :
\begin{align}
\widehat t(\ve{\ell}) f_{\widehat\vR}(\vk,t)\widehat t^{-1}(\ve{\ell}) = f_{\widehat\vR}(\vk,t). 
\label{eq:fMagTransl}
\end{align}
The above condition puts constraints on both the ground state distribution function $f_{0,\widehat{\vR}}$ and the coadjoint orbit deformation $\phi_{\widehat{\vR}}$.

{For $f_{0,\widehat\vR}(\vk)$, combining \eqref{eq:35} and \eqref{eq:fMagTransl}, we have}
\be
\bra{\ve{K}}f_{0,\widehat\vR}(\vk)\ket{\ve K'} =  \tilde{f}_{0,\ve K}(\vk)\delta_{\ve{K}\ve{K}'}.
\label{eq:f0Mag}
\ee
As we consider a fixed chemical potential, $\tilde{f}_{0,\ve K}$ does not depend on the eigenvalue $\ve{K}$ explicitly. In the semiclassical regime, $\vk$ remains a continuous variable, so the ground state distribution function is $f_{0,\widehat \vR}(\vk)=\tilde{f}_{0,\ve K}(\vk) = f_0 (\vk, t)=\Theta(k_F-|\vk|)$\footnote{Fixing the chemical potential as $B$ increases from zero, the Fermi wave vector should satisfy $p_F=k_F$, where $p_F$ is defined in Eq.~\eqref{eq:f0B0}.}. The subgroup of canonical transformations $\mathcal{H}$ that leave $f_0$ invariant (Eq.~\eqref{eq:stablizer}) should be defined accordingly with the distribution function in Eq.~\eqref{eq:f0Mag}.
Using \eqref{eq:fHybrid}, Eq.~\eqref{eq:fMagTransl} gives
\begin{align}
\widehat t(\ve{\ell}) \phi_{\widehat\vR}(\vk,t)\widehat t^{-1}(\ve{\ell}) &= \phi_{\widehat\vR}(\vk,t).
\label{eq:38}
\end{align}
Following the same argument, using \eqref{eq:35}, we have
\be
\bra{\ve{K}}\phi_{\widehat\vR}\ket{\ve K'} = \tilde\phi_{\ve K}\delta_{\ve{K}\ve{K}'}.
\ee

Importantly, we see that as a matrix in the $\ket{\ve{K}}$ basis, even though $\phi_{\widehat\vR}$ has $N_{\Phi}^2$ independent components, magnetic translation symmetry constrains the number of pertinent modes down to the $N_{\Phi}$. {If we proceeded without this constraint, we would retain all $N_{\Phi}^2$ modes and ultimately obtain an incorrect free energy that is not extensive, which indicates some modes are overcounted. This is related to the UV-IR mixing issue in bosonization without $B$ field.~\cite{DDMS2022}.}

Carrying out the trace, the bosonized action is
\be
S =\sum_{\ve{K}\in {\rm mBZ}} \int \diff t  \[\Langle f_{0}(\vk), \iu\,\tilde U_{\ve K}^{-1}\star_\vk \partial_t \tilde U_{\ve{K}} \Rangle-\Langle \tilde f_{\ve{K}},\epsilon(\vk)\Rangle\],
\label{eq:41}
\ee
where $\tilde U_{\ve{K}} =\exp(\iu \tilde \phi_{\ve{K}})$, {$\tilde f_{\ve{K}} = \tilde U_{\ve{K}}\star_\vk f_{0} \star_\vk \tilde U^{-1}_{\ve{K}}$, and as we discussed, the Moyal product (in $\vk$) can be expanded and truncated at leading nontrivial order}. 
As an expansion, the action can be written in terms the compact boson 
\be
\tilde\phi_{\ve{K}}(\vk, t) \simeq \tilde\phi_{\ve{K}}(\vk, t)+ 2\pi
\ee
as $S=S_{\rm WZW}+ S_{H}$ with
\begin{widetext}
\begin{align}
S_{\rm WZW}[\phi] =& \sum_{\ve{K}\in {\rm mBZ}}\int \diff t \Langle f_0, \(-\dot\phi_{\ve{K}} - \frac{1}{2} \{\phi_{\ve{K}}, \dot{\phi}_{\ve{K}}\} - \frac{1}{6 }\{\phi_{\ve{K}}, \{\phi_{\ve{K}}, \dot{\phi}_{\ve{K}}\}\} + \mathcal{O}(\phi^4)  \)  \Rangle \label{eq:43}\\
S_{H}[\phi] =& \sum_{\ve{K}\in {\rm mBZ}} \int \diff t \Langle  \(f_0-\{\phi_{\ve{K}},f_0\} +\frac{1}{2} \{\phi_{\ve{K}}, \{\phi_{\ve{K}}, f_0\}\} - \frac{1}{6} \{\phi_{\ve{K}}, \{\phi_{\ve{K}}, \{\phi_{\ve{K}},f_0\}\}\} + \mathcal{O}(\phi^4)  \), -\epsilon(\vk)  \Rangle,\non
\end{align}
\end{widetext}
where we have made the replacement $\tilde\phi\to\phi$ for convenience. We remind that $\{,\}$ and $\langle,\rangle$ are the Poisson bracket (Eq.~\eqref{eq:poissonk}) and the inner product (Eq.~\eqref{eq:innerk}) for $\vk$.

At this step, it is common (see e.g., Ref~\cite{DDMS2022}) to ``flip the Poisson bracket"  in $S_{\rm WZW}$ using $
\Langle f, \{F, \phi_{\ve{K}}\} \Rangle = - \Langle \{f,\phi_{\ve{K}}\}, F  \Rangle$,
which seems to follow straightforwardly from integration by parts. However, while boundary terms indeed vanish, the correct full expression should be
\be
\!\!\!\Langle f, \{F, \phi_{\ve{K}}\} \Rangle = - \Langle \{f,\phi_{\ve{K}}\}, F  \Rangle + B\Langle f, F\,\nabla\times\nabla\phi_{\ve{K}}\Rangle.
\label{eq:44}
\ee
For single-valued analytic functions, $\nabla\times\nabla\phi_{\ve{K}}=0$ obviously and the last term is redundant, but it is not the case when $\phi_{\ve{K}}$, being a angular variable, has a winding configuration $\phi_{\ve{K}}\sim \tilde{p}\theta$. In the following, we will first derive the theory without winding modes of $\phi_{\ve{K}}$ (either in $t$ or in $\theta$), and then return to the proper treatment of these modes. {In Sec.~\ref{sec:quant}, we show that the inclusion of these modes are crucial in the quantization of the bosonized theory.}

\subsection{Chiral bosons in momentum space\label{sec:ChiralBosonAction}}

For the remainder of this work, we focus on the bosonized action for non-interacting fermions with quadratic dispersion $\epsilon(\bm k) = \bm k^2/2m - \mu$. Without including winding configurations, we can flip the Poisson brackets with $\phi_{\ve{K}}$ at each order in both lines of Eq.~\eqref{eq:43}, such that on the co-adjoint side of the inner product we have
\be
\{ f_0, \phi_{\ve{K}}\}=B\delta(|\vk|-k_F)\partial_{\theta}\phi_{\ve{K}}(\vk)/k_F,
\ee
where $\tan\theta=k_y/k_x$.
Due to the $\delta$-function, the $\vk$ integral can be reduced to an angular integral over $\theta$, and the $\phi_{\ve {K}}$ fields are fixed to $|\vk|=k_F$. Without loss of generality, we take $\phi_{\ve {K}}$ near the FS to be almost a constant in $k$ and depend only on $\theta$.\footnote{From the mode expansion \eqref{eq:mode} in Sec.~\ref{sec:quant}, $\phi_{\ve{K}}(\vk)$ cannot be exactly a constant in $k$, and $\partial_{k}^n\phi_{\ve{K}}(\vk)$ at large enough $n$ will become significant. However, these only occur in Moyal corrections, which are irrelevant perturbations guaranteed by Eq.~\eqref{eq:25}.} Remarkably, cubic and higher-order terms in both $S_{\rm WZW}$ and $S_{H}$ vanish, since 
\begin{align}
\{\phi_{\ve{K}}(\theta),\dot\phi_{\ve{K}}(\theta)\}=&0 \\
\{\phi_{\ve{K}}(\theta),\{\phi_{\ve{K}}(\theta),\epsilon(\ve k)\}\} =&\{\phi_{\ve{K}}(\theta),\omega_c \partial_\theta\phi_{\ve{K}}(\theta)\} =0,\non
\end{align}where in the second line we have used for parabolic dispersion, the cyclotron frequency
\be
\omega_c(|\vk|) \equiv \frac{B\partial_{k}\epsilon(\vk)}{k_F} = \frac{B}{m}
\ee
is independent of $|\vk|$. Therefore, the semiclassical bosonized EFT in a homogeneous magnetic field becomes {Gaussian}.  This drastically differs from the bosonized theory for a generic FS without a magnetic field, in which the theory is necessarily {nonlinear}~\cite{DDMS2022} with infinite irrelevant interaction terms. We emphasize that the noninteracting nature of the EFT is only transparent in the $(\vR, \vk)$ coordinates; in the $(\vx, \vk)$ basis~\cite{Huang2024}, the theory is perturbatively connected to that without a magnetic field, and retains all the interaction terms.
We also emphasize that the free theory is obtained only for a parabolic dispersion; for a generic dispersion, there are higher-order nonlinear terms in $S_H$, the effects of which we leave to a future study.

After some simplification  the bosonized action reads, up to a constant,
\begin{align}
\!\!\!\!S[\phi]=& \sum_{i=1}^{N_{\Phi}}\left\{\int \frac{\diff t \diff \theta}{4\pi}  
\[\partial_\theta\phi_{i}
\,\dot\phi_{i} - {\omega_c} \(\partial_\theta\phi_{i}\)^2\] + S_{\rm w}^i\right\}.
\label{eq:FJ}
\end{align}
where to simplify notation the summation over $\ve{K}$ in the mBZ is relabeled with summation over $i$ from 1 to $N_{\Phi}$. The terms in $S_{\rm w}$ {are integrals of total derivatives, and} are nonzero only when $\phi_i(\theta,t)$ has a winding configuration, either in $\theta$ or $t$, which we discuss in the next Section.

As the theory splits into decoupled term with identical actions, $N_{\Phi}=BL_1L_2/2\pi$ has the clear meaning of LL degeneracy, which emerges naturally from bosonizing the FS in a magnetic field.

\begin{figure}
\includegraphics[width=\columnwidth]{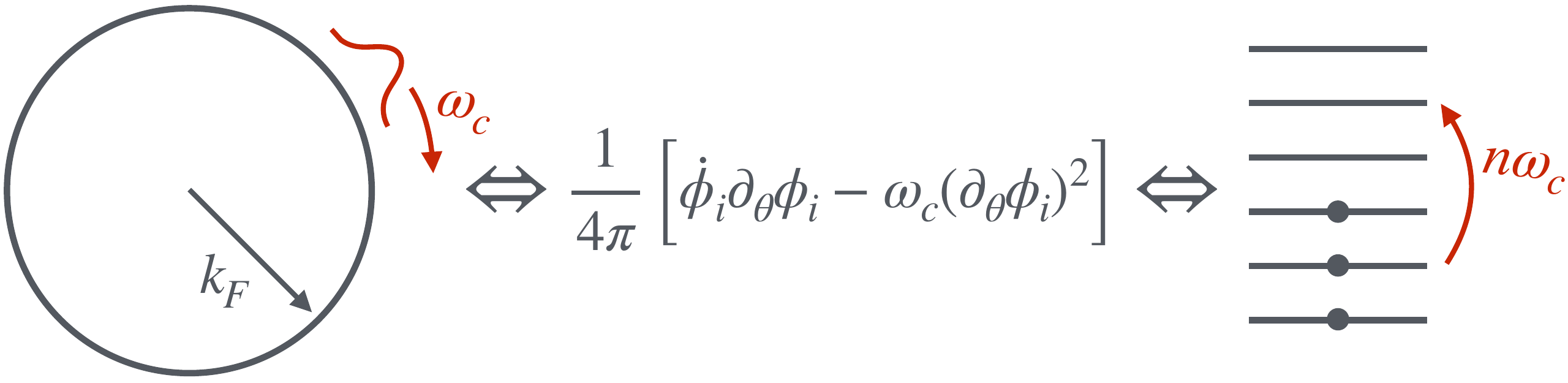}
\caption{Illustration of the bosonized theory of a 2d FS in a weak magnetic field, in which the cyclotron motion around the FS is described by a chiral boson action. The energy quanta of this action are found to correspond to the particle-hole excitations among LLs (see Sec.~\ref{sec:quant}). }
\label{fig:cyclo}
\end{figure}

Apart from $S_{\rm w}$, Eq.~\eqref{eq:FJ} is a Floreanini-Jackiw action~\cite{FJ1987} describing to $N_{\Phi}$ species of chiral bosons in momentum space with an angular speed $\omega_c$, which we illustrate in Fig.~\ref{fig:cyclo}.  Similar actions have been obtained in the $(\vx,\vp)$ basis in earlier works~\cite{Fradkin2018,Huang2024}, but our results have two key differences. First, in the current basis the action is much simpler, and as we will see, the $N_{\Phi}$ decoupled bosons lead directly to the LL degeneracy. Second, in our theory there exist an additional contribution $S_{\rm w}$ from the winding configurations that was neglected, which again, as we shall see, has an important physical meaning. The chiral boson picture was implied (without the explicit momentum-space action) in an earlier work~\cite{Caldeira1997} by directly bosonizing the excitation between LLs. However, our theory does not assume \emph{a priori} the LLs; instead, as we shall see the LL physics is naturally reproduced from the bosonic description of the Fermi surface in a magnetic field.

The chiral boson action in position space describes the edge state of an integer quantum Hall state~\cite{Wen1995} with a Chern number equal to $N_{\Phi}$. This precisely matches the semiclassical picture electron wave packets moving along the FS.  From this perspective, the cyclotron motion can be interpreted as edge states of a Chern insulator in momentum space. We note that this analogy has been recently obtained via a ``bulk" phase-space Chern-Simons action~\cite{Else2021PRX,ma2021emergent,else2023holographic,Yizhuang2024,Hughes2024,kunyang-2025}
\be
S=\frac{1}{24\pi^2}\int_{k_x,k_y,x,y,t} \mathcal{A}\wedge \diff \mathcal{A}\wedge \diff\mathcal{A},
\ee
describing the LU(1)-anomaly for Fermi surfaces first proposed in Ref.~\cite{Else2021PRX}. To see this, note that in our case the gauge field has a background value $(d\mathcal A)_{x,y}= B$, and the action becomes
\be
S=\frac{N_{\Phi}}{4\pi}\int_{k_x,k_y,t} \mathcal{A}\wedge \diff \mathcal{A},
\ee
which perfectly matches the chiral anomaly of the action \eqref{eq:FJ}.

From our derivation, it is clear that this analogy  only holds in the semiclassical limit $\omega_c\ll \mu$, which we used to derive Eq.~\eqref{eq:43}. Indeed, In the opposite limit the system should be instead regarded as a Chern insulator in \emph{real space} with a Chern number equal to the number of occupied LLs.

\section{Quantization of the bosonized action}
\label{sec:quant}
In this section we quantize the bosonized action \eqref{eq:FJ} for free fermions with a parabolic dispersion. We obtain the full spectrum of the bosonic Hamiltonian, which we show can be  matched to that of the fermionic theory. Alternatively, one can also directly compute the thermodynamic functions from the path integral (partition function), which is needed in the next section.

To simplify notations, let us focus on a single flavor of the boson $\phi_i \equiv \phi$ with 
\begin{align}
S= S_{\rm cb}+S_{\rm w},~~
S_{\rm cb}=&\int \frac{\diff t \diff \theta}{4\pi} \[\partial_\theta\phi\dot\phi - {\omega_c} \(\partial_\theta\phi\)^2\]
\label{eq:FJ2}
\end{align}
and the additional piece $S_{\rm w}$ is given by, according to \eqref{eq:43} and \eqref{eq:44},
\begin{widetext}
\begin{align}
S_{\rm w} =& -\int \diff t \[\Langle f_0, \dot\phi\Rangle 
-\Langle f_0, B\(\nabla\times\nabla\phi\)\(\frac{1}{2}\dot\phi+\frac16\{\phi,\dot\phi\}+...\)\Rangle \right. \non\\
&\left.+\Langle B\(\nabla\times\nabla\phi\)\(\frac{1}{2}\{\phi,f_0\} - \frac{1}{6}\{\phi,\{\phi,f_0\} \}+...\),\epsilon(\vk)\Rangle\].
\end{align}
\end{widetext}

Naively neglecting the issue of winding modes and $S_{\rm w}$, canonical quantization of this action leads to the Kac-Moody algebra~\cite{Fradkin2018,Else2021PRX} in $\theta$ direction
\be
\[\hat\rho(\theta),\hat\rho(\theta')\] = -\frac{i}{2\pi}\partial_{\theta}\delta(\theta - \theta'),
\label{eq:kacmoody}
\ee
where 
\be
\hat\rho(\theta) \equiv -\frac{\partial_{\theta}\hat\phi(\theta)}{2\pi},
\label{eq:density}
\ee
which according to Eq.~\eqref{eq:fHybrid} is the density of electrons (compared with equilibrium value)  with momentum $\vk=(k_F\cos\theta, k_F\sin\theta)$. Indeed, choosing a gauge in which $\phi(\vk)=\phi (\theta)$~\cite{DDMS2022}, we have
\begin{align}
\int \frac{\diff \vk}{2\pi B} \[f(\vk)-f_0(\vk)\] &=  -\int \frac{\diff \vk}{2\pi B} B\partial_\theta \phi\, \delta(k-k_F) +\cdots\non\\
&= -\int d\theta \frac{\partial_\theta\phi}{2\pi},
\label{eq:density2}
\end{align}
where the terms in $\cdots$ contain derivatives of $\delta(k-k_F)$, which vanishes upon integration.
This commutation relation has appeared in Refs.~\cite{Fradkin2018,Else2021PRX}, and used in Ref.~\cite{Else2021PRX} for an elegant non-perturbative proof of the Luttinger's theorem in 2d.

From Eq.~\eqref{eq:density2}, a grand canonical ensemble of fermions can be achieved by {extending the target manifold of the bosonized field thery and incorporate} singular, winding configurations of $\phi$ such that $\int d\theta \phi(\theta) \neq 0$.
In the presence of  winding modes the quantization of the action is  more subtle. To this end, it is convenient to express the $\phi(\theta)$ field at the FS via a mode expansion
\begin{align}
\phi(\theta) = {q} + \tilde p \theta + \sum_{n\neq 0}\frac{a_n}{\sqrt{|n|}}  \eu^{\iu n\theta},~~~\theta\in[-\pi, \pi),
\label{eq:mode}
\end{align}
where {$a^*_n=a_{-n}$,} $q\simeq q+2\pi$, and $\tilde p$ is a winding number.

In order to evaluate $S_{\rm w}$, we can analytically continue $\phi(\theta)$ to inside the Fermi sea, the exact form of which amounts to a gauge choice.~\cite{DDMS2022} This can be accomplished by requiring 
\be
a_n(\vk ,t)\propto k^{|n|}\lambda(t)
\ee
{near $\vk=0$} {(but asymptotically becomes a constant in $k$ near the FS)}. In addition, importantly, any nonzero $\tilde p$ leads to a $\vk$-space vortex configuration with a singularity at $\vk =0$, and thus 
\be
\nabla\times\nabla \phi = 2\pi \tilde{p}\, \delta(\vk).
\ee
With this expression of $\phi(\vk)$, it is straightforward to show that only two terms survive in $S_{\rm w}$, i.e.,
\be
S_{\rm w} = -\int \diff t \Langle f_0, \dot\phi - \frac{B}{2}\(\nabla\times\nabla \phi\)\dot\phi\Rangle.
\label{eq:sw}
\ee

{By substituting Eq.~\eqref{eq:mode} into Eqs.~(\ref{eq:FJ2}) and (\ref{eq:sw}), the contributions to the action from zero (non-oscillatory) modes $(\tilde p, q)$ and oscillatory modes $\{a_n\}$ split as 
\be
S[\phi]=S_{\rm Fock}[\{a_n\}] + S_{\rm zero}[\tilde p, q].
\ee
For the oscillatory modes $a_n$, $S_{\rm w}=0$, and $S_{\rm Fock}$ comes solely from the chiral boson action. Canonically quantizing it, we have the standard} commutation relation
\be
\[\hat a_{-n},\hat a_m\]=\delta_{nm}\sgn(m),
\label{eq:60}
\ee
and the corresponding Hamiltonian is found to be 
\be
\hat H_{\rm Fock} = \omega_c\sum_{n=1}^{\infty}n \(\hat a_{-n}\hat a_n + \frac{1}{2}\).
\ee
Aside from the zero-point energy, this Hamiltonian for one of the $N_{\rm \Phi}$ bosonic species corresponds precisely to the transitions among LLs,~\cite{Caldeira1997,Fradkin2018,Du-Mehta-Son-2021}, with $n$ being the difference of LLs. We illustrate this correspondence in Fig.~\ref{fig:cyclo}. {We note that one additional subtlety is that operators such as $\hat a_n$ and $\hat a_{1}^n$ carry the same energy, and should correspond to different LL transitions of the same magnitude. It is indeed possible to directly establish a one-to-one correspondence with the Fock states on the fermionic side while keeping the particle-hole transformation intact~\cite{ye_wang_unpublished}. We will not do so here, but will indirectly show this in the next Section by matching the bosonic and fermionic partition functions. }

{For the zero modes $(\tilde p, q)$, both $S_{\rm cb}$ and $S_{\rm w}$  contribute to $S_{\rm zero}$.} We have
\be
S_{\rm zero} [\tilde p,q] = \int \diff t \(-\frac{\mu}{\omega_c} \dot q + \tilde p\dot q - \frac{\omega_c}{2}\tilde p^2\).
\ee
This is precisely the action for a particle moving on a ring with a unit radius ($q\simeq q+2\pi$), in the presence of a magnetic flux penetrating the center of the ring. This fictitious magnetic flux, not to be confused with the actual magnetic flux, is equal to $-2\pi\mu/\omega_c$. {Canonically quantizing $S_{\rm zero}$ we get $[\hat q,\hat{\tilde p}]=\iu$. {It is worth noting that this commutation relation is necessary to reproduce the Kac-Moody algebra \eqref{eq:kacmoody}, which can be seen from Eqs.~(\ref{eq:mode}) and (\ref{eq:60}).}

{To obtain the spectrum of $\hat{\tilde p}$ and the Hamiltonian one needs to perform the path integral.} As is well-known, the first term in $S_{\rm zero} [\tilde p,q]$ is a topological $\theta$-term in $0+1$d~\cite{altland}. {The path integral in the zero mode sector in imaginary time with $it\to \tau\in [0,\beta)$ is given by
\begin{align}
Z_{\rm zero} 
=&\int \mathcal{D}\tilde p(\tau) \mathcal{D}q(\tau)  \exp[\left. -\iu\(\frac{\mu}{\omega_c}- \tilde p\)q\right|_0^\beta] \\
&\times \exp[\int_0^\beta d\tau \,\iu q\frac{d}{d\tau}\(\frac{\mu}{\omega_c}- \tilde p\) - \frac{\omega_c}{2}\tilde p^2], \non
\end{align}
where we have integrated by parts in $\tau$. Integrating out $q(\tau)$  under $q(\beta)=q(0) \mod 2\pi$ leads to the constraint that $\tilde p-\mu/\omega_c$ must be a constant \emph{and} an integer. The result is
\begin{align}
Z_{\rm zero} =\sum_{p\in\mathbb{Z}} \exp[-\frac{\beta\omega_c}{2}\(p+\frac{\mu}{\omega_c}\)^2].
\label{eq:62}
\end{align}
We can immediately read off the corresponding Hamiltonian as
\be
\hat H_{\rm zero} = \frac{\omega_c}{2}\[\hat p+\Delta\(\frac{\mu}{\omega_c}\)\]^2,~~~\mathrm{spec}(\hat p)=\mathbb Z,
\ee
where
\be
\Delta(x)\equiv x- \lfloor x +1/2  \rfloor \in [-1/2,1/2)
\ee
denotes the distance between $x$ and its nearest integer, and we have absorbed this integer for $\mu/\omega_c$ into $p$.} Mapped to the fermion side, this zero-mode energy also has a clear physical meaning: it is precisely the energy cost of adding/substracting $p$ electrons {into the Fermi sea with fixed chemical potential.}

Subtracting the vacuum energy, the normal-ordered Hamiltonian incorportating all excitations from $N_{\Phi}$ channels is 
\begin{align}
\!\!\!\normord{\hat H} \,=& {\normord{\hat H_{\rm Fock} + \hat H_{\rm zero}}}\nonumber\\
=&\omega_c \sum_{i}^{N_{\Phi}} \left\{\[\sum_{n=1}^{\infty}n \hat a^i_{-n}\hat a_n^i \] + \frac{\hat p_i^2}{2}+\Delta\(\frac{\mu }{\omega_c}\)\hat p_i\right\}.
\label{eq:spec}
\end{align}

We see that as $\mu$ varies, the energy spectrum displays oscillatory behavior that is periodic in $\mu/\omega_c$. This is the origin of the dHvA effect, as we shall see in detail in the next Section. We remind that in our theory the oscillatory behavior originates from the the WZW term of the action, which after mode expansion includes a topological $\theta$-term. Being a topological effect, such an oscillation cannot be captured in any perturbative expansion in $B$. 

\section{Thermal and magnetic responses}
\label{sec:resp}

{In this Section, we first present a  proof of the equivalence between the bosonic and fermionic theories by relating their partition functions.  By this proof, our bosonized action is capable of capture all thermal and magnetic properties of a Fermi surface under a weak magnetic field. Instead of stopping there, we also obtain these properties directly from the bosonic theory, which will be useful for future studies of Fermi liquids and non-Fermi liquids via a bosonization approach.}

\subsection{Matching bosonic and fermionic partition functions}
From the full bosonic spectrum in Eq.~\eqref{eq:spec} {(or directly performing the path integral in imaginary time)},  we obtain the partition function  of the bosonic theory at a temperature $T=1/\beta$ as 
\be
Z= Z_0 \Tr[\exp(-\beta \normord{\hat H})]\equiv Z_0 Z_T,
\label{eq:zt}
\ee
where $Z_0=\exp(-\beta E_0)$ with $E_0$ being the ground state energy, which cannot be computed within the EFT, and $Z_T$ is the contribution from thermal excitations. $Z_T$ can be factorized into two parts. First, the spectrum of the oscillaltory sector is that of a massless boson in 1d. The second contribution comes from the zero modes. Explicitly, we have
\be
Z=Z_0 \(\sum_{{p}=-\infty}^{\infty}\eu^{-{\beta\omega_c} {p}^2/2-\beta\omega_c {p} \Delta } \prod_{n=1}^{\infty}\frac{1}{1-\eu^{-n\beta\omega_c}}\)^{N_\Phi},
\label{eq:zb}
\ee
where we used the shorthand $\Delta\equiv \Delta(\mu/\omega_c)$. {To ensure a typical eigenstate remains in the semiclassical regime, we take $T\ll \mu$.}

Interestingly, $Z$ can be reexpressed by making use of Jacobi triple product identity,
\begin{widetext}
\begin{align}
\prod_{n=1}^\infty
\left( 1 - x^{2n}\right)
\left( 1 + x^{2n-1} y^2\right)&\left( 1 +\frac{x^{2n-1}}{y^2}\right)
=\sum_{p=-\infty}^\infty x^{p^2} y^{2p},
\end{align}
which holds for $|x|<1$. Setting $x=\exp(-\beta\omega_c/2)$ and $y=\exp(-\beta\omega_c\Delta/2)$,
we have
\begin{align}
Z=&Z_0\[\prod_{n=1}^{\infty}\(1+\eu^{-\beta\omega_c(n-1/2+\Delta)}\)\(1+\eu^{-\beta\omega_c(n-1/2-\Delta)}\)\]^{N_{\Phi}}.
\end{align}
\end{widetext}
This is precisely the fermionic (grand) partition function in the {$\mu\gg\omega_c, T$ limit (such that the summation over hole states can be extended to infinity)}, where $Z_0$ is interpreted as contribution from the Fermi sea, {i.e.\ the LLs below $\mu$}, and the two factors at each $n$ correspond to particle and hole excitations, and $\omega_c\Delta(\mu/\omega_c)$ is the offset of chemical potential from the middle of two neighboring LLs. 

{We have thus established that the bosonized action Eq.~\eqref{eq:FJ} fully describes the 2d fermionic system in a weak magnetic field. We note that a simillar relation holds in 1d bosonization~\cite{haldane1981} obtained from the Hamiltonian formalism, although the counterpart of $\Delta(\mu/\omega_c)$ is absent. }

\subsection{Responses from the bosonic theory}

All response functions can be computed from the free energy $F=-T\log Z$. At first sight, it seems impossible to obtain dHvA effect from Eq.~\eqref{eq:zb}, as we know from the fermionic theory that at low $T$ it mainly comes from the oscillation of the ground state energy $E_0(B)$, which is incorportated in the unknown piece $Z_0$. Rather, we can only compute the contributions from thermal excitations, $Z_T$ (see Eq.~\eqref{eq:zt}).

To proceed, we use an additional physical input that dHvA effect must vanish at $\mu\gg T\gg \omega_c$. The free energy is expressed as
\be
F= -T\log Z_0 -T\log Z_T \equiv  E_0(B)  + F_T.
\ee
Since the contribution to the free energy from $Z_0$ is just $E_0$, which is temperature independent, this means that $F_T$ at $T \gg \omega_c$ must contain a $T$-independent oscillatory piece that is exactly opposite to the dHvA oscillation of $E_0$, which we can compute. Aside from this piece, $F_T$ also contains the dHvA oscillations at finite temperatures, as well as terms accounting for Landau diamagnetism and heat capacity.

Explicitly, from Eq.~\eqref{eq:zb} we have
\begin{align}
F_T =&N_{\Phi}T \sum_{n=1}^{\infty}\log(1-\eu^{n\beta\omega_c}) \label{eq:67} \\
 &-N_{\Phi}T \log\(\sum_{{p}=-\infty}^{\infty} \eu^{-{\beta\omega_c} ({p}+\Delta)^2/2}\) -N_{\Phi}\frac{\omega_c\Delta^2}2. \non
\end{align}

The last term of \eqref{eq:67} is the $T$-independent oscillatory piece {that persists at $T\gg \omega_c$}, 
which means the ground state energy 
\be
E_0(B) - E_0(0) = \frac{N_{\Phi}\omega_c}2 \[\Delta\(\frac{\mu}{\omega_c}\)\]^2 + \textrm{non-osc.},
\label{eq:69}
\ee
where ``non-osc." denotes non-oscillatory terms. $E_0$ oscillates with a period of 
\be
\delta\(\frac{1}{B}\) = \frac{1}{m\mu} =\frac{2\pi}{\mathcal{A}_{\rm FS}}.
\ee
Oscillation with the same period exists in the zero-temperature magnetic moment $M=-\partial E_0/\partial B$, which is the dHvA effect. Setting non-oscillatory terms to zero, Eq.~\eqref{eq:69} precisely matches the result from the fermionic side with a parabolic dispersion.\footnote{The non-oscillatory terms cannot be obtained within the bosonized theory. For a generic dispersion, it may be nonzero. {We leave the question of separating the temperature independent nonoscillatory contributions from $F_0$ and $F_T$ for future study.}}  We plot Eq.~\eqref{eq:69} in Fig.~\ref{fig:dHvA} (in blue), which is indeed identical to the result from the fermionic theory.

The free energy can thus be written as
\begin{align}
F= &N_{\Phi}T \sum_{n=1}^{\infty}\log(1-\eu^{-n\beta\omega_c})\non\\
&-N_{\Phi}T \log\(\sum_{p=-\infty}^{\infty} \eu^{-{\beta\omega_c} (p+\Delta)^2/2}\)+ E_0(B=0).
\label{eq:71}
\end{align}
There remains an oscillatory contribution, which comes from the second term. Using the Poisson resummation formula for the second term, we obtain 
\begin{widetext}
\begin{align}
F = & N_{\Phi}T \sum_{n=1}^{\infty}\log(1-\eu^{-n\beta\omega_c})-N_{\Phi}T\log\(\int_{-\infty}^{\infty} dx\, \eu^{-{\beta\omega_c} x^2/2}+2\sum_{k=1}^\infty \int_{-\infty}^{\infty} dx\,\cos(2\pi k x)\eu^{-{\beta\omega_c} (x+\Delta)^2/2}\) + E_0(B=0) \non\\
=& \underset{F_{\rm no}}{\underbrace{\rule[-14pt]{0pt}{0pt} N_{\Phi}T \sum_{n=1}^{\infty}\log(1-\eu^{-n\beta\omega_c})-\frac{N_{\Phi}T}2\log\(\frac{2\pi T}{\omega_c}\)+ E_0(B=0)}}-N_\Phi T \log\[1+2\sum_{k=1}^\infty \cos\(2\pi k \frac{\mu}{\omega_c}\)\eu^{-2\pi^2k^2T/\omega_c}\]\non\\
\approx & F_{\rm no}-2N_{\Phi} T \eu^{-2\pi^2T/\omega_c}\cos\(2\pi \frac{\mu}{\omega_c}\),
\label{eq:poisson}
\end{align}
\end{widetext}
where $F_{\rm no}$ are the nonoscillatory terms. In the last line we have only kept the $k=1$ term and expanded the log, which holds  for $T\gtrsim \omega_c$ at leading order in $\eu^{-T/\omega_c}$.
Compared with that in $E_0$, the oscillation in $F$ has the same period, but is expontially supressed, with a temperature-dependent amplitude
\be
A(T)= -2TN_{\Phi}\exp(\frac{-2\pi^2 T}{\omega_c}).
\ee
Not surprisingly, this agrees with the result from the fermionic side, known as the Lifshitz-Kosevich formula
\be
A_k(T)= \frac{(-)^k TN_{\Phi}}{k\sinh({2k\pi^2 T}/{\omega_c})},
\label{eq:LK}
\ee
for $k=1$ and $T\gg \omega_c$. Directly obtaining Eq.~\eqref{eq:LK} from Eq.~\eqref{eq:poisson} requires extra work~\cite{ye_wang_second}, but it is guaranteed since, as we showed, the bosonic and fermionic partition functions match.

At $T\gg \omega_c$, the predominant contributions to response function are non-oscillatory, which can be obtained by taking the derivative over non-oscillatory parts of the free energy $F_{\rm no}$. In practice, it turns out to be slightly simpler to compute them from the total energy $E=\partial (\beta F_{\rm no})/\partial \beta$, given by (after subtracting the constant piece $E_0(B=0)$)
\begin{align}
E=& N_{\Phi}\sum_{n=1}^{\infty}\frac{n\omega_c}{\eu^{n\beta\omega_c}-1}+\frac{N_{\Phi}T}2  \non\\
=&N_{\Phi}\sum_{n=0}^{\infty}\frac{n\omega_c}{\eu^{n\beta\omega_c}-1}-\frac{N_{\Phi}T}2 .
\end{align}
In the second line, the value of the summand at $n=0$ is defined through the $n\to 0$ limit. Using the Euler-MacLaurin formula, and neglecting nonanalytic, exponentially suppressed terms in $B$,  we obtain
\begin{align}
E=& N_{\Phi}\int \frac{\diff x x\omega_c}{\eu^{x\beta\omega_c}-1} +\frac{N_\Phi \omega_c}{24}=\frac{\pi^2 g(\mu)T^2}{6} + \frac{\mu_B^2g(\mu) B^2}{6},
\label{eq:76}
\end{align}
where $g(\mu)=mL^2/2\pi$ is the fermionic density of states, $\mu_B=1/2m$ is the Bohr magneton. This result resembles that for black-body radiation in 1d.

From the first term of \eqref{eq:76}, we obtain the heat capacity 
\be
C=\frac{\pi^2 g(\mu)T}{3},
\ee
which is precisely the result from the fermionic theory. (Note that for a 2d electron gas with parabolic dispersion, the heat capacity does not receive any corrections at higher order in $T$.) We note that in the absence of magnetic field, obtaining the correct coefficient for the specific heat was found to be challenging using bosonization~\cite{DDMS2022}  due to UV-IR mixing. From our result, the magnetic field can be viewed as a UV regulator with $\ell_B$ as an effective short-distance cutoff. 

The second term of \eqref{eq:76} gives rise to Landau diamagnetism:
\be
\chi_{\rm dia} = -\frac{\partial^2(E-TS)}{\partial B^2} = -\frac{1}{3}\mu_B^2 g(\mu).
\ee
In obtaining this result, we have used the fact that $S=\int C(T)dT/T$, which in our case is $B$-independent. {The same result can also be directly obtained from $\chi_{\rm dia}=-\partial^2 F_{\rm no}/\partial B^2$, which can be shown straightforwardly by first taking the second-order derivative in $B$ for the summand, and then perform the summation $\sum_{n=1}^{\infty}$ using the Euler-MacLaurin formula. }

Combining with the oscillatory terms in Eq.~\eqref{eq:poisson}, we get the $B$-dependence of the free energy as
\be
F(B) - F(0) = -N_{\Phi}T\eu^{-2\pi^2T/\omega_c}\cos(2\pi\frac{\mu}{\omega_c})+\frac{\mu_B^2}{6}g(\mu)B^2.
\label{eq:80}
\ee
Eq.~\eqref{eq:80} is plotted in Fig.~\ref{fig:dHvA} (in orange), which agrees with the result from the fermionic theory.


\begin{figure}
\includegraphics[width=\columnwidth]{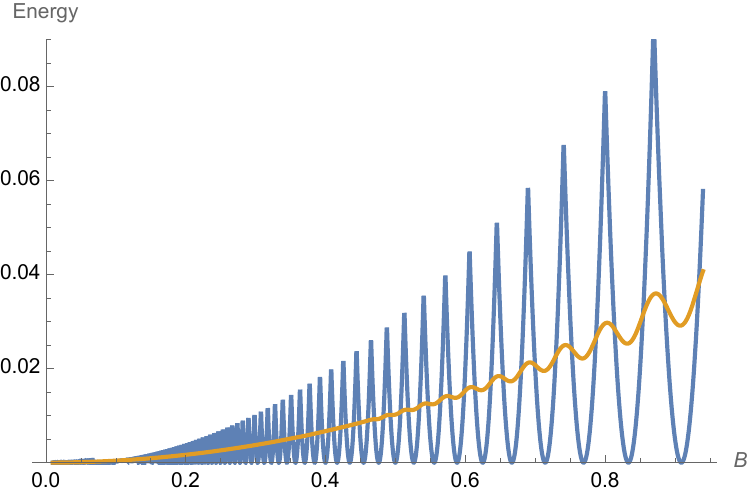}
\caption{The dependence of the ground state energy (blue) and the (finite-temperature) free energy (orange) on the external magnetic field $B$. We have set the density of states $g(\mu)=1$, the electron mass $m=1$, chemical potential $\mu=10$, and temperature $T=0.15$.}
\label{fig:dHvA}
\end{figure}

\section{Conclusion and outlook}
\label{sec:conc}
In this work, we extended the coadjoint-orbit method for nonlinear bosonization of the Fermi surface to systems under a weak magnetic field. By properly treating the noncommutativeness of the base manifold, we showed that the effective field theory is that for $N_{\Phi}$ flavors of chiral bosons in momentum space. {This theory can be interpreted as the edge states a quantum Hall insulator in momentum space, which can also be viewed as a consequence of the LU(1) 't Hooft anomaly associated with a Fermi surface}. {For free electron gas with parabolic dispersion, the energy spectrum of the chiral bosons are integer times of the cyclotron frequency $\omega_c$, which physically corresponds to the LL transitions.} {Furthermore, the action contains total derivative terms {which were not included in the literature. Specifically we uncovered a term} linear in bosonized field, which upon mode expansion becomes a topological $\theta$-term.} After quantizing the theory, we obtained the thermal and magnetic responses of a 2d Fermi surface.  Notably we showed that the dHvA effect, which is inherently a nonperturbative effect and challenging to obtain via field-theoretic approaches, has a topological origin in the bosonic theory. Our approach thus reveals the connections between LL quantization, the phase of a semiclassical cyclotron orbit, and the emergent LU(1) 't Hooft anomaly associated with a Fermi surface.

{The EFT approach developed here points to a few interesting directions to study a generic Fermi surface in a weak magnetic field.}

\emph{Generalizing to FLs and NFLs.} Bosonization is a powerful approach to describe the low-energy physics of gapless fermions in the presence of interaction effects. Notably it offers an alternative starting point to describe the behavior of non-Fermi liquids. For example, in Ref.~\cite{DDMS2022} it was found by coupling to a critical boson with $\vq=0$, the theory naturally leads to a $\sim T^{2/3}$ specific heat, {which is a NFL contribution}.
{The bosonization method we developed in this work can be extended to study the magnetic properties,}
{including the thermodynamic potential~\cite{Nosov2024} and collective modes~\cite{Fradkin2018} in a magnetic field,} of strongly correlated gapless fermionic systems. To this end, one open question is to write down the fermion interaction terms in the EFT using the $(\vR, \vk)$ basis. Whether the interaction terms are Landau parameters or Yukawa couplings to a critical boson, they are local in the original spatial coordinates $\vx$, and need to be properly expressed and analyzed in the $(\ve{R},\ve{k})$ basis. We leave this issue to future work~\cite{wang2025FL}.

\emph{Incorporating disorder effects.} Disorder effects leads to a ``Dingle factor" that reduces the amplitudes of the dHvA effect. Moreover, they are crucial for transport properties such as the Shubnikov-de Haas effect. A microscopic way of describing disorder effect is to add random fermion bilinears that breaks magnetic translation symmetry. In the current framework, these terms are $\sim V \cos(\phi_{i}-\phi_{j})$, where $\langle V\rangle=0$ and $\langle V^2\rangle\neq 0$. On the other hand, on a phenomenological level, the finite lifetime effects can be described by a Lorentzian broadening of the chemical potential. A detailed analysis of disorder at both levels is an important open problem.

\emph{Generalizing to Fermi surfaces with non-trivial Berry phases.}In this work, we have focused a FS in a single band system, which does not have any nonzero Berry connection from the Bloch wave functions. The bosonized action could also be extended to a FS realized in a multi-band system, where the geometric properties of the Bloch wave function, e.g.\ the Berry curvature, can lead to interesting consequences in the magnetic properties. While earlier studies constructed in phase space suggested that the Berry connection corresponds to momentum-space gauge fields in the semiclassical limit $\nabla_{\vx}\cdot\nabla_{\vp} \ll 1$~\cite{Chang_2008}, it would be beneficial to understand the effect of nontrivial Bloch wave functions systematically in Moyal algebra, and incorporate Berry phase effects to a FS with or without a magnetic field. Furthermore, previous studies~\cite{Mikitik1999,Alexandradinata2023ar} of Berry phases in dHvA effect largely relied on the semiclassical picture, and our results provide a framework to construct an effective field theory in which correlation effects can be naturally incorporated, which will be addressed in a subsequent work~\cite{ye_wang_second}. 

\acknowledgments
 We would like to thank Leon Balents, Andrey Chubukov, Luca V.\ Delacr\'etaz, Yi-Hsien Du, Dominic Else, Eduardo Fradkin, Leonid Glazman, Xiaoyang Huang, Johannes Knolle, Umang Mehta, Srinivas Raghu, Inti Sodemann, Dam T.\ Son, and Xiao-Chuan Wu  for useful discussions. YW is supported by NSF under award number DMR-2045781. MY is supported by a start-up grant from the University of Utah. We acknowledge support by grant NSF PHY-1748958 to the Kavli Institute for Theoretical Physics (YW), and by grant NSF PHY-2210452 to the Aspen Center for Physics (MY), where this work was partly performed.

\appendix

\section{Derivation of bosonized action from coherent-state path integral}
\label{app:path}
In this Appendix, we derive in details the bosonized action Eq.~\eqref{eq:11} for free fermions via a path integral approach. {The coherent-state path integral derivation was first presented in the bosonization of a patched Fermi surface in Ref.~\cite{FradkinBosonization1994}.
The derivation here has been sketched in several talks~\cite{Luca_talk,Umang_talk}. We thank the authors for illuminating discussions.}

Denoting the ground state with a filled Fermi sea up to the chemical potential as $\ket{\rm FS}$, a coherent state in the many-body Hilbert space is given by~\cite{MehtaThesis2023}
\be
\ket{\mathcal{U}} = \hat{\mathcal{U}}\ket{\rm FS},
\label{eq:a0}
\ee
where $\hat{\mathcal{U}} = \exp (\iu \phi_{ij} c_i^\dagger c_j)$ (with repeating indices summed) is generated by fermion bilinears, and $i,j\in [1,L^2]$ can be viewed as site indices (or lattice momentum indices). 

Via the algebra of fermion bilinears, it is straightforward to show that the all possible $\hat{\mathcal{U}}$'s form a Lie group. {For the Hamiltonian dynamics of free fermions, it can be shown that the time-evolution operator $\mathrm{exp}(\iu \hat H \diff t)$ commutes with $\int \diff \,\mathcal{U} |\mathcal{U}\rangle\langle\mathcal{U}|$, where  $\diff\,\mathcal{U}$ is the Haar measure~\cite{altland} of the Lie group, and thus by Schur's Lemma, up to an unimportant prefactor,
\be
\int \diff \,\mathcal{U} |\mathcal{U}\rangle\langle\mathcal{U}|=1.
\label{eq:a1}
\ee
If $\hat H$ has additional symmetries, one may further restrict the  $\hat{\mathcal{U}}$'s to a subgroup that preserves the same symmetry, as we do in  Sec.~\ref{sec:constraint}. }

Much like the path integral for spins~\cite{altland}, by inserting an resolution of identity operators  via coherent states \eqref{eq:a1} at every infinitesimal time interval,  the path integral can be written as
\be
\!\!\!\! Z = \int \mathcal D \mathcal{U}(t) \exp[\int dt \bra{\rm FS} \hat{\mathcal{U}}^{-1}\(-\partial_t-\iu \hat H\)\hat{\mathcal{U}}\ket{\rm FS}],
\label{eq:Haar}
\ee
where  $\hat H$ is the Hamiltonian.


We note that $\hat{\mathcal{U}}^{-1}(\partial_t-\iu \hat H)\hat{\mathcal{U}}$ is by definition expressed as a sum of nested commutators of fermion bilinear operators of the form $\hat \Omega = c^\dag_i \omega_{ij} c_j$. Even though they are operators in a $2^N$-dimensional Hilbert space, they can be represented by much smaller, $N\times N$ matrices $\omega_{ij}$, which are operators in the first-quantized Hilbert space. Namely, we can prove that from fermion statistics, for $\hat A = c^\dag_i a_{ij} c_j$ and $\hat B = c^\dag_i b_{ij} c_j$,
\be
[\hat A, \hat B] = c_i^\dag (a_{ik} b_{kj} - b_{ik} a_{kj})c_j = c_i^\dag ([\widehat a, \widehat b])_{ij}c_j,
\label{eq:a3}
\ee
where the first-quantized operators, denoted by lower-case letters and a wider hat symbol,  are defined through $\bra{i}\widehat a\ket{j}\equiv a_{ij}$. 

By expanding $\hat{\mathcal{U}}$, using \eqref{eq:a3}, and then re-exponentiating the nested first-quantized commutators, we can now rewrite the action as
\be
iS = \int \diff t \bra{\rm FS} c^\dag_i c_j\ket{\rm FS}\[\widehat U^{-1} (-\partial_t-\iu \widehat \epsilon\,) \widehat U\]_{ij},
\ee
where $\widehat U$ and $\widehat \epsilon$ are first-quantized operators: 
\begin{align}
\widehat U\equiv \eu^{\iu \widehat \phi},
\end{align}
with $\bra{i}\widehat{\phi}\ket{j}\equiv\phi_{ij}$, and $\widehat\epsilon$ is the one-particle Hamiltonian, $\hat H = c^\dagger_i \epsilon_{ij} c_j$.

We can define the one-particle density matrix (also known as the ``statistical matrix") operator $\widehat f_0$ for the ground state via its elements~\cite{Land5}
\be
f_{0,ji} \equiv \bra{\rm FS}c^\dag_i c_j \ket{\rm FS},
\ee
such that, e.g., the ground-state total energy can be rewritten as $E_0=\bra{\rm FS}\hat H \ket{\rm FS} =\Tr(\widehat f \,\widehat \epsilon)$. We then obtain
\begin{align}
S =& \int \diff t \Tr[\widehat f_0 \widehat U^{-1}\(\iu\partial_t - \widehat \epsilon\,\)\widehat U] \nonumber\\
=& \int \diff t \Tr[\widehat f_0 \widehat U^{-1}\iu\partial_t\widehat U -\widehat f \,\widehat \epsilon]
\label{eq:a7}
\end{align}
where $\widehat f \equiv \widehat U \widehat f_0\widehat U^{-1}$, whose elements can be shown to satisfy $f_{ji}=\bra{\Psi}c^\dag_i c_j\ket{\Psi}$, i.e., form the one-particle density matrix of the coherent state. We remind that the trace is done in the $N$-dimensional first-quantized Hilbert space, and all operators are one-particle operators.

The Wigner function is introduced via
\begin{align}
\omega(\vx,\vp) =\int d\ve{y}\,e^{\iu \vp\cdot \ve y/\hbar}\bra{\vx-\frac{\ve{y}}2}\widehat\omega\ket{\vx + \frac{\ve{y}}2}.
\end{align}
It practice, for an operator explicitly expressed as $\omega(\widehat\vx,\widehat\vp)$, its Wigner function is  $\omega(\vx,\vp)$. The Wigner function  satisfies two useful properties. First, the trace of the product of operators is converted to a phase-space integral:
\begin{align}
\Tr(\widehat a\, \widehat b) &=\int\frac{d\vx d\vp}{(2\pi\hbar)^2} a(\vx,\vp)\star b(\vx, \vp)
\label{eq:a8}
\end{align}
Second, the Wigner function of the commutator of two operators can be expressed via the Moyal bracket and the star product [see Eqs.~(\ref{eq:1},\ref{eq:2})] of their Wigner functions, i.e., for $\widehat c = [\widehat a, \widehat b\,]$,
\begin{align}
c(\vx, \vp) =& \iu\,\{a(\vx, \vp),b(\vx, \vp)\}_{\rm M} \nonumber\\
=& a(\vx, \vp)\star b(\vx, \vp) - b(\vx, \vp)\star a(\vx, \vp).
\end{align}

Using these two properties, one can rewrite the action \eqref{eq:a7} in terms of the Wigner function as
\begin{align}
S 
=\int \diff t \int\frac{d\vx d\vp}{(2\pi\hbar)^2} &\[f_0(\vx,\vp)
U^{-1}(\vx,\vp,t)\star\iu\partial_t U(\vx,\vp,t) \right. \nonumber\\
&\left.- f(\vx,\vp) \epsilon(\vp)\]
\end{align}
where $U(\vx,\vp)=\exp\(\iu \phi(\vx,\vp)\)$ is the Wigner function of $\widehat U$, $f(\vx, \vp)=U(\vx, \vp) \star f_0(\vp) \star U^{-1}(\vx, \vp)$ is the Wigner function of $\widehat{f}$, and $\epsilon(\vp)$ is the Wigner function of $\epsilon(\widehat\vp)$.
This is the action \eqref{eq:11} presented in Sec.~\ref{sec:IIB} for noninteracting fermions.

\bibliography{MagBosonization}

\end{document}